\documentclass{ws-procs975x65}
\usepackage{theorem,amsmath,amssymb,mathrsfs}
\usepackage[hypertex,linkcolor=red]{hyperref}
\def\index#1{}
\def\<{\langle}\def\>{\rangle}
\def\:{\hbox{\bf :}}
\renewcommand{\leq}{\leqslant}

\def\Reals{\mathbb R}\def\Cmplx{\mathbb C}
\def\map#1{{\mathscr{#1}}}

\def\set#1{{\sf #1}}\def\alg#1{{\mathcal #1}}\def\aA{\alg{A}}
\def\aI{\alg{I}}

\def\dim{\operatorname{dim}}\def\adm{\operatorname{dim}}
\def\idim#1{\operatorname{dim}_\#(#1)}

\def\Span{\set{Span}}

\def\sH{\set{H}}

\def\ie{i. e. }
\def\n#1{|\!|#1|\!|}
\def\dag{\dagger}\def\eff{{\rm eff}}
 
\newtheorem{postulate}{Postulate}

\def\trnsfrm#1{\mathscr #1}
\def\tA{\trnsfrm A}\def\tB{\trnsfrm B}\def\tC{\trnsfrm C}
\def\tS{\trnsfrm S}
\def\tI{\trnsfrm I}\def\tT{\trnsfrm T}\def\tU{\trnsfrm U}\def\tX{\trnsfrm X}

\def\AA{\mathbb A}\def\AB{\mathbb B}\def\AL{\mathbb L}
\def\cA{{\underline{\tA}}}\def\cB{\underline{\tB}}\def\cC{\underline{\tC}}\def\cT{\underline{\tT}}

\def\Stset{{\mathfrak S}}\def\Wset{{\mathfrak W}}
\def\Trnset{{\mathfrak T}}
\def\Cntset{{\mathfrak P}}\def\Ball{{\mathfrak B}_1}

\begin{document}

\title{Where the mathematical structure of Quantum Mechanics comes from
}

\author{Giacomo Mauro D'Ariano$^*$}
\address{Dipartimento di Fisica "A. Volta" dell'Universit\`a di Pavia, via Bassi 6, 27100 Pavia, Italy\\
$^*$E-mail: dariano@unipv.it,  webpage: www.qubit.it\\
Dept. of Electrical and Computer Engineering, Northwestern University, Evanston, IL 60208}
\begin{abstract}
  The mathematical formulation of Quantum Mechanics is derived from purely operational axioms based
  on a general definition of {\em experiment} as a set of transformations. The main ingredient of
  the mathematical construction is the postulated existence of {\em faithful states} that allows one
  to calibrate the experimental apparatus. Such notion is at the basis of the operational
  definitions of the scalar product and of the {\em adjoint} of a transformation.
\end{abstract}
\bodymatter
\section{Introduction}
In spite of its unprecedented predicting power in the whole physical domain, the starting point of
Quantum Mechanics is purely mathematical, with no direct physical interpretation of the formalism.
Undeniably Quantum Mechanics is not based on a set of physical laws or principles from which the
mathematical framework is derived---as we would expect from a theory. Considering the universality
of Quantum Mechanics, its "physical" axioms should be of very general nature, transcending Physics
itself, at the higher epistemological level, and should be related to {\em observability principles}
that must be satisfied independently on the specific physical laws object of the experiment. In
previous works \cite{darianoVax2005,dariano-losini2005,darianoVax2006} I showed how it is possible
to derive the Hilbert space formulation of Quantum Mechanics from five operational Postulates
concerning {\em experimental accessibility and simplicity}. In the present paper I will give a
synthetical presentation of this axiomatization: additional details and mathematical proofs can be
found in Ref.~\refcite{darianoVax2006}.  The mathematical formulation of Quantum Mechanics in terms
of complex Hilbert space for finite dimensions is derived starting from the five Postulates. For the
infinite dimensional case a C${}^*$-algebra representation of physical transformations is derived
from only four of the five Postulates, via a Gelfand-Naimark-Segal (GNS)
construction\cite{GelfandNeumark}. The starting point for the axiomatization is a seminal definition
of {\em physical experiment}, which, as first shown in Ref. ~\refcite{darianoVax2005}, entails a
thorough series of notions that lie at the basis of the axiomatization. The postulated existence of a
{\em faithful state}, which allows one to calibrate the experimental apparatus, provides operational
definitions for the scalar product and for the {\em adjoint} of a transformation at the core of the
C${}^*$-algebra representation of transformations via the Gelfand-Naimark-Segal (GNS) construction.
This crucial ingredient from the present axiomatization comes from modern Quantum Tomography
\cite{tomo_lecture}, and concerns the possibility of performing a complete quantum calibration of
measuring apparatuses \cite{calib} and transformations \cite{tomo_channel} by using a single pure
bipartite state \cite{faithful}.

\section{The postulates} 
The general background is that in any experimental science we make {\em experiments} to get {\em
  information} on the {\em state} of an {\em object physical system}. Knowledge of such a state will
allow us to predict the results of forthcoming experiments on the same object system. Since we
necessarily work with only partial {\em a priori} knowledge of both system and experimental
apparatus, the rules for the experiment must be given in a probabilistic setting.

\paragraph{General Axiom: On what is an experiment.}\label{ga:2} An experiment on an
  object system consists in making it interact with an apparatus. The interaction between object
  and apparatus produces one of a set of possible transformations of the object, each one occurring
  with some probability. Information on the ``state'' of the object system at the beginning of the
  experiment is gained from the knowledge of which transformation occurred, which is the "outcome"
  of the experiment signaled by the apparatus.
\begin{postulate}[Independent systems]\label{p:independent} There exist independent physical systems.
\end{postulate}
\begin{postulate}[Informationally complete observable]\label{p:infocom} For each physical system
  there exists an informationally complete observable. 
\end{postulate}
\index{local observability principle} 
\begin{postulate}[Local observability principle]\label{p:locobs} For every composite system there exist
  informationally complete observables made only of local informationally complete observables.
\end{postulate}
\begin{postulate}[Informationally complete discriminating observable]\label{p:Bell} For every system
  there exists a minimal informationally complete observable that can be achieved using a joint
  discriminating observable on the system + an ancilla (i.e. an identical independent system). 
\end{postulate}
\begin{postulate}[Symmetric faithful state]\label{p:faith} For every composite system made of two identical
  physical systems there exist a symmetric joint state that is both dynamically and preparationally faithful. 
\end{postulate}
\medskip

\section{The statistical and dynamical structure}

\par According to our definition of {\bf experiment}---the starting point of our
axiomatization---the experiment is identified with the set $\AA\equiv\{\tA_j\}$ of possible
transformations $\tA_j$ that can occur on the object system. The apparatus will signal the {\bf
  outcome} $j$ labeling which transformation actually occurred. The experimenter cannot control
which transformation occurs, but he can decide which experiment to perform, namely he can choose the
set of possible transformations $\AA=\{\tA_j\}$. For example, in an Alice\&Bob communication
scenario Alice will encode the different bit values by choosing between two experiments
$\AA=\{\tA_j\}$ and $\AB=\{\tA_j\}$ corresponding to two different sets of transformations
$\{\tA_j\}$ and $=\{\tB_j\}$.  The experimenter has control on the transformation itself only in the
special case when the transformation $\tA$ is deterministic. In the following, wherever we consider
a nondeterministic transformation $\tA$ by itself, we always regard it in the context of an
experiment, namely assuming that there always exists at least a complementary transformation $\tB$
such that the overall probability of $\tA$ and $\tB$ is unit.
\par Now, since the knowledge of the state of a physical system allows us to predict the results of
forthcoming possible experiments on the system (more generally, on another system in the same
physical situation), namely it would allow us to evaluate the probabilities of any possible
transformation for any possible experiment, then, by definition, a {\bf state} $\omega$ for a
physical system is a rule that provides the probability for any possible transformation, namely
$\omega$ is a state means that $\omega(\tA)$ is the probability that the transformation $\tA$
occurs.  We clearly have the completeness condition $\sum_{\tA_j\in\AA}\omega(\tA_j)=1$, and we will
assume that the identical transformation $\tI$ occurs with probability one, \ie $\omega(\tI)=1$,
corresponding to a special choice of the lab reference frame as in the {\em Dirac picture}. In the
following for a given physical system we will denote by $\Stset$ the set of all possible states and
by $\Trnset$ the set of all possible transformations. In order to include also non-disturbing
experiments, we must conceive situations in which all states are left invariant by each
transformation. It is convenient to extend the notion of state to that of {\bf weight}, \ie a
nonnegative bounded functionals $\tilde\omega$ over the set of transformations with
$0\leq\tilde\omega(\tA)\leq\tilde\omega(\tI)<+\infty$ for all transformations $\tA$.  To each weight
$\tilde\omega$ it corresponds the properly normalized state $\omega=\tilde\omega/\omega(\tI)$.
Weights make the convex cone $\Wset$ generated by the convex set of states $\Stset$.
\par When composing two transformations $\tA$ and $\tB$, the probability $p(\tB|\tA)$ that $\tB$
occurs conditional on the previous occurrence of $\tA$ is given by the Bayes rule for conditional
probabilities $p(\tB|\tA)=\omega(\tB\circ\tA)/\omega(\tA)$.  This sets a new probability rule
corresponding to the notion of {\bf conditional state} $\omega_\tA$ which gives the probability that
a transformation $\tB$ occurs knowing that the transformation $\tA$ has occurred on the physical
system in the state $\omega$, namely $\omega_\tA\doteq\omega(\cdot\circ\tA)/\omega(\tA)$ (in the
following we will make extensive use of the functional notation with the central dot corresponding
to a variable transformation). One can see that the present definition of ``state'', which logically
follows from the definition of experiment, leads to the identification {\em
  state-evolution}$\equiv${\em state-conditioning}, entailing a {\em linear action of
  transformations on states} (apart from normalization) $\tA\omega:=\omega(\cdot\circ\tA)$: this is
the same concept of {\bf operation} that we have in Quantum Mechanics, giving the conditional state
as $\omega_\tA=\tA\omega/\tA\omega(\tI)$. In other words, this is the analogous of the
Schr\"{o}dinger picture evolution of states in Quantum Mechanics. One can see that in the present
context linearity of evolution is just a consequence of the fact that the evolution of states is
pure state-conditioning: this will include also the deterministic case
$\tU\omega=\omega(\cdot\circ\tU)$ of transformations $\tU$ with $\omega(\tU)=1$ for all states
$\omega$---the analogous of quantum unitary evolutions and channels.

From the Bayes conditioning it follows that we can define two complementary types of equivalences
for transformations: the {\em dynamical} and {\em informational} equivalences. The transformations
$\tA_1$ and $\tA_2$ are {\em dynamically equivalent} when $\omega_{\tA_1}=\omega_{\tA_2}$
$\forall\omega\in\Stset$, whereas they are {\em informationally equivalent} when
$\omega(\tA_1)=\omega(\tA_2)$ $\forall\omega\in\Stset$. The two transformations are then completely
equivalent when they are both dynamically and informationally equivalent, corresponding to the
identity $\omega(\tB\circ\tA_1)=\omega(\tB\circ\tA_2)$,
$\forall\omega\in\Stset,\;\forall\tB\in\Trnset$. We call {\bf effect} an informational equivalence
class of transformations (this is the same notion introduced by Ludwig\cite{Ludwig-axI}).  In the
following we will denote effects with the underlined symbols $\cA$, $\cB$, etc., or as $[\tA]_\eff$,
and we will write $\tA_0\in\cA$ meaning that "the transformation $\tA$ belongs to the equivalence
class $\cA$", or "$\tA_0$ corresponds to the effect $\cA$'', or "$\tA_0$ is informationally
equivalent to $\tA$".  Since, by definition one has $\omega(\tA)\equiv\omega(\cA)$, we will
legitimately write $\omega(\cA)$ instead of $\omega(\tA)$. Similarly, one has $\omega_\tA(\tB)\equiv
\omega_\tA(\cB)$, which implies that $\omega(\tB\circ\tA)=\omega(\cB\circ\tA)$, which gives the
chaining rule $\cB\circ\tA\in\underline{\tB\circ\tA}$ corresponding to the "Heisenberg picture"
evolution of transformations acting on effects (notice that in this way transformations act from the
right on effects). Now, by definitions effects are linear functionals over states with range
$[0,1]$, and, by duality, we have a convex structure over effects.  We will denote the convex set of
effects by $\Cntset$.

\par The fact that we necessarily work in the presence of partial knowledge about both object and
apparatus corresponds to the possibility of incomplete specification of both states and
transformations, entailing the convex structure on states and the addition rule for {\em coexistent
transformations}, namely for transformations $\tA_1$ and $\tA_2$ for which 
$\omega(\tA_1)+\omega(\tA_2)\leq 1,\;\forall\omega\in\Stset$ (\ie transformations that can in
principle occur in the same experiment). The addition of the two coexistent transformations is the
transformation $\tS=\tA_1+\tA_2$ corresponding to the event   $e=\{1,2\}$ in which the apparatus
signals that either $\tA_1$ or $\tA_2$ occurred, but does not specify which one. Such transformation
is specified by the informational and dynamical equivalence classes $\forall\omega\in\Stset$:
$\omega(\tA_1+\tA_2)=\omega(\tA_1)+\omega(\tA_2)$ and
$(\tA_1+\tA_2)\omega=\tA_1\omega+\tA_2\omega$. Clearly the composition "$\circ$" of transformations
is distributive with respect to the addition "$+$". We will also denote by
$\tS(\AA):=\sum_{\tA_j\in\AA} \tA_j$ the deterministic transformation $\tS(\AA)$ corresponding to
the sum of all possible transformations $\tA_j$ in $\AA$. We can also define the multiplication
$\lambda\tA$ of a transformation $\tA$ by a scalar $0\leq\lambda\leq 1$ as the transformation which
is  dynamically equivalent to $\tA$, but occurs with rescaled probability
$\omega(\lambda\tA)=\lambda\omega(\tA)$. Now, since for every couple of transformation
$\tA$ and $\tB$ the transformations $\lambda\tA$  and $(1-\lambda)\tB$ are coexistent for
$0\leq\lambda\leq 1$, the set of transformations also becomes a convex set. Moreover, since the
composition $\tA\circ\tB$ of two transformations $\tA$ and $\tB$ is itself a transformation and
there exists the identical transformation $\tI$ satisfying $\tI\circ\tA=\tA\circ\tI=\tA$ for every
transformation $\tA$, the transformations make a semigroup with identity, \ie a {\em monoid}.
Therefore, the set of physical transformations is a convex monoid.

It is obvious that we can extend the notions of coexistence, sum and multiplication by a scalar from
transformations to effects via equivalence classes.

\par A purely dynamical notion of {\bf independent systems} coincides with the possibility of
performing local experiments. More precisely, we say that two physical systems are {\em independent}
if on the two systems 1 and 2 we can perform {\em local experiments} $\AA^{(1)}$ and $\AA^{(2)}$
whose transformations commute each other (\ie
$\tA^{(1)}\circ\tB^{(2)}=\tB^{(2)}\circ\tA^{(1)},\;\forall \tA^{(1)}\in\AA^{(1)},\,\forall
\tB^{(2)}\in\AB^{(2)}$).  Notice that the above definition of independent systems is purely
dynamical, in the sense that it does not contain any statistical requirement, such as the existence
of factorized states. Indeed, the present notion of dynamical independence is so minimal that it can
be satisfied not only by the quantum tensor product, but also by the quantum direct sum. As we will
see in the following, it is the local observability principle of Postulate \ref{p:locobs} which will
select the tensor product. In the following, when dealing with more than one independent system, we
will denote local transformations as ordered strings of transformations as follows
$\tA,\tB,\tC,\ldots:=\tA^{(1)}\circ\tB^{(2)}\circ\tC^{(3)}\circ\ldots$. For effects one has the
locality rule $([\tA]_\eff,[\tB_\eff)\in[(\tA,\tB)]_\eff$. The notion of independent
systems now entails the notion of {\em local state}---the equivalent of partial trace in Quantum
Mechanics. In the presence of many independent systems in a joint state $\Omega$, we define the {\bf
  local state} $\Omega|_n$ of the $n$-th system as the probability rule
$\Omega|_n(\tA)\doteq\Omega(\tI,\ldots,\tI,\underbrace{\tA}_{n\text{th}},\tI,\ldots)$ of the joint
state $\Omega$ with a local transformation $\tA$ only on the $n$-th system and with all other
systems untouched. For example, for two systems we write $\Omega|_1=\Omega(\cdot,\tI)$.
\par We conclude this section by noticing that our definition of dynamical independence implies the
acausality of correlations between independent systems---the so-called {\em no-signaling}---\ie: Any
local "action" (\ie experiment) on a system does not affect another independent system. In
equations: $\forall\Omega\in\Stset^{\times 2},\forall\AA$, $\Omega_{\tS(\AA),\tI}|_2=\Omega|_2$.
Notice that even though the no-signaling holds, the occurrence of the transformation $\tB$ on system
1 generally affects the local state on system 2, i. e. $\Omega_{\tB,\tI}|_2\neq\Omega_2$, and such
correlations can be checked {\em a posteriori}. We emphasize that the no-signaling is a mere
consequence of our minimal notion of dynamical independence.

\section{Banach structure}
We can extend the convex cone of weights to its embedding linear space by taking differences of
weights, and forming {\em generalized weights}. We will denote the linear space of generalized
weights as $\Wset_\Reals$. Likewise we can extend effects and transformations to generalized effects
and transformations, whose linear spaces will be denoted by $\Cntset_\Reals$ and $\Trnset_\Reals$,
respectively. The linear space $\Trnset_\Reals$ of generalized transformations inherits a real
algebra structure from the convex monoid of physical transformations $\Trnset$.  On the linear
spaces $\Wset_\Reals$, $\Cntset_\Reals$, and $\Trnset_\Reals$ we can now superimpose a Banach space
structure, by introducing norms in form of supremum. We start from physical effects for which we
define the norm as the supremum of the respective probability over all possible physical states. We
then extend the norm to generalized effects $\cA\in\Cntset_\Reals$ by taking the absolute value, \ie
$\n{\cA}:=\sup_{\omega\in\Stset}|\omega(\cA)|$. It is easy to check that this is indeed a norm. We
can now introduce the unit ball $\Ball:=\{\cA\in\Cntset_\Reals,\, \n{\cA}\leq 1\}$ and define the
norm for weights as $\n{\tilde\omega}:=\sup_{\cA\in\Ball}|\tilde\omega(\cA)|$.  For transformations
we then introduce the norm in the standard way used for linear operators over Banach spaces, namely
$\n{\tA}:=\sup_{\cB\in\Ball}\n{\cB\circ\tA}$, which is equivalent to the double supremum
$\n{\tA}=\sup_{\cB\in\Ball}\sup_{\omega\in\Stset}|\omega(\cB\circ\tA)|$. It is then easy to check
that $\Trnset_\Reals$ becomes a real Banach algebra (\ie it satisfies the norm inequality
$\n{\tB\circ\tA}\leq\n{\tB}\n{\tA}$). It is crucial to perform the supremum over the unit ball,
instead of just physical effects: this guarantees the Banach algebra structure for generalized
transformations.  It is also clear that physical transformation correspond to contractions, \ie they
have bounded norm $\n{\tA}\leq1$, whence the convex monoid of physical transformations $\Trnset$ has
the form of a truncated convex cone. As a corollary, we have that two physical transformations $\tA$
and $\tB$ are coexistent iff $\tA+\tB$ is a contraction. We also have the bound between
transformation and effect norms $\n{\cA}\leq\n{\tA}$, with the identity for $\tA$ in the double
cone. Operationally all norm closures correspond to assume preparability (of effects, states, and
transformations) by an approximation criterion in-probability. The norm closure may not be required
operationally, however, as any other kind of extension, it is mathematically very convenient. The
convex set of states $\Stset$ and the convex sets of effects $\Cntset$ are dual each other under the
pairing $\omega(\cA)$ giving the probability of effect $\cA$ in the state $\omega$. Therefore, the
convex set of effects is a truncated convex cone of positive linear contractions over the convex set
of states, namely the set of bounded positive functionals $0\leq l\leq 1$ on $\Stset$, with
$l_{\cA}(\omega):=\omega(\cA)$. Such duality can be trivially extended to generalized effects and
generalized weights via the pairing $|\omega(\cA)|$, and $\Wset_\Reals$ and $\Cntset_\Reals$ become
a dual Banach pair. This Banach space duality is the analogous of the duality between bounded
operators and trace-class operators in Quantum Mechanics.  It is worth noticing that this dual
Banach pair is just a consequence of the probabilistic structure that is inherent in our starting
definition of experiment.

\section{Observables}
The {\em observable} is just a complete set of effects $\AL=\{l_i\}$ of an experiment
$\AA=\{\tA_j\}$, namely one has $l_i=\underline{\tA_j}$ $\forall j$. Clearly, one has the
completeness relation $\sum_il_i=1$. The observable $\AL=\{l_i\}$ is {\em informationally complete}
when each effect $l$ can be written as a linear combination $l=\sum_ic_i(l)l_i.$ of elements of
$\AL$, or, in other words, $\Cntset_\Reals\equiv\Span(\AL)$. We will call the informationally
complete observable {\em minimal} when its effects are linearly independent. Clearly, using an
informationally complete observable we can reconstruct any state $\omega$ from just the
probabilities $l_i(\omega)$ as $\omega(\cA)=\sum_ic_i(l_{\underline{\tA}})l_i(\omega)$: this is just
the Bloch representation of states. In such representation the Banach structure manifests itself in
a vector representation for states and effects, and in a matrix representation for transformations,
the physical transformations corresponding to affine linear maps.

We will call an effect (and likewise a transformation) $\cA$ {\em predictable} if there exists a
state for which $\cA$ occurs with certainty and another state for which it never occurs, and {\em
  resolved} if there is only a single pure state for which it occurs with certainty. Similarly an
experiment will be called {\em predictable} when it is made only of predictable effects, and {\em
  resolved} when all its effects are resolved. For a predictable effect $\cA$ one has $\n{\cA}=1$,
and for a predictable transformation $\tA$ one has $\n{\tA}=1$.  Notice that a predictable
transformation is not necessarily deterministic. Predictable effects $\cA$ correspond to affine
functions $f_\tA$ on the state space $\Stset$ with $0\leq f_\tA\leq 1$ achieving both bounds. We
call a set of states $\{\omega_n\}_{n=1,N}$ {\em perfectly discriminable} if there exists a
predictable and resolved experiment $\AL=\{l_j\}_{j=1,N}$ which discriminates the states, \ie
$\omega_m(l_n)=\delta_{nm}$. We call {\em informational dimension} of the convex set of states
$\Stset$, denoted by $\idim{\Stset}$, the maximal cardinality of perfectly discriminable set of
states in $\Stset$. Clearly, an observable $\AL=\{l_j\}$ is {\em discriminating} and {\em resolved}
for $\Stset$ when $|\AL|\equiv\idim{\Stset}$, \ie $\AL$ discriminates a maximal set of discriminable
states.

We now come to the notions of faithful state. We say that a state $\Phi$ of a composite system is
{\em dynamically faithful} for the $n$th component system when for every transformation $\tA$ the
map $\tA\leftrightarrow(\tI,\ldots,\tI,\underbrace{\tA}_{n\text{th}},\tI,\ldots)\Phi$ is one-to-one,
with the transformation $\tA$ acting locally only on the $n$th component system. Physically, the
definition corresponds to say that the output conditioned weight (\ie the conditioned state
multiplied by the probability of occurrence) is in one-to-one correspondence with the
transformation.  Restricting attention to bipartite systems, a state is dynamically faithful (for
system 1) when $(\tA,\tI)\Phi=0\;\Longleftrightarrow\tA=0$, which means that for every bipartite
effect $\cB$ one has $\Phi(\cB\circ(\tA,\tI))=0\quad\Longleftrightarrow\quad\tA=0$.  Clearly the
correspondence remains one-to-one when extended to $\Trnset_\Reals$. On the other hand, we will call
a state $\Phi$ of a bipartite system {\em preparationally faithful} for system 1 if every joint
bipartite state $\Omega$ can be achieved by a suitable local transformation $\tT_\Omega$ on system 1
occurring with nonzero probability. Clearly a bipartite state $\Phi$ that is preparationally
faithful for system 1 is also locally preparationally faithful for system 1, namely every local
state $\omega$ of system 2 can be achieved by a suitable local transformation $\tT_\omega$ on system 1.
\par In Postulate \ref{p:faith} we also use the notion of {\em symmetric} joint state. This is
simply defined as a joint state of two identical systems such that for any couple of transformations
$\tA$ and $\tB$ one has $\Phi(\tA,\tB)=\Phi(\tB,\tA)$.

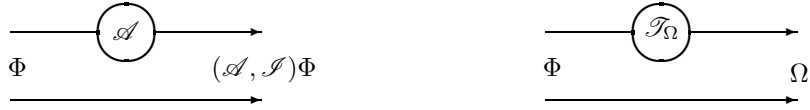
\begin{figure}[h]
\begin{center}
\parbox{2.1in}{
\setlength{\unitlength}{800sp}
    \begin{picture}(8745,3219)(931,-3565)
      {\thicklines \put(5401,-1261){\oval(1756,1756)}}
      {\put(1801,-1261){\line(1, 0){2700}}}
      {\put(6301,-1261){\vector(1, 0){3300}}}
      {\put(1801,-3361){\vector(1, 0){7800}}}
      \put(2026,-2611){\makebox(0,0)[b]{$\Phi$}}
      \put(5401,-1486){\makebox(0,0)[b]{$\tA$}}
      \put(9676,-2811){\makebox(0,0)[b]{$(\tA,\tI)\Phi$}}
    \end{picture}
}
  \hspace*{44pt}
  \parbox{2.1in}{
    \setlength{\unitlength}{800sp}
    \begin{picture}(8745,3219)(931,-3565)
      {\thicklines \put(5401,-1261){\oval(1756,1756)}}
      {\put(1801,-1261){\line(1, 0){2700}}}
      {\put(6301,-1261){\vector(1, 0){3300}}}
      {\put(1801,-3361){\vector(1, 0){7800}}}
      \put(2026,-2611){\makebox(0,0)[b]{$\Phi$}}
      \put(5401,-1486){\makebox(0,0)[b]{$\tT_\Omega$}}
      \put(9676,-2811){\makebox(0,0)[b]{$\Omega$}}
    \end{picture}
}
  \caption{{\bf Left:} Illustration of the notion of {\em dynamically faithful} state for a bipartite
    system. The state $\Phi$ is dynamically faithful when the output weight (conditioned state
    multiplied by the probability of occurrence) is in one-to-one correspondence with the
    transformation. {\bf Right:} Illustration of the notion of preparationally faithful state for a
    bipartite system. The state $\Phi$ is {\em preparationally faithful} for system 1 if every joint
    bipartite state $\Omega$ can be achieved by a suitable local transformation $\tT_\Omega$ on
    system 1 occurring with nonzero probability. 
}
\end{center}
\end{figure}

\section{Dimensionality theorems}\label{s:finiteHilbert}
We now consider the consequences of Postulates \ref{p:locobs} and \ref{p:Bell}. The {\em local
  observability principle} (Postulate \ref{p:locobs}) is operationally crucial, since it reduces
enormously the experimental complexity, by guaranteeing that only local (although jointly executed)
experiments are sufficient to retrieve a complete information of a composite system, including all
correlations between the components. The principle reconciles holism with reductionism, in the sense
that we can observe an holistic nature in a reductionistic way---\ie locally.  This principle
implies identity ($D_3$) in Table \ref{t:dimids} for the affine dimension of the convex set of a
bipartite systems as a function of the dimensions of the components. This identity is the same that
one obtains in Quantum Mechanics due to the tensor product structure.  We conclude that the tensor
product is not a consequence of dynamical independence in Def.  \ref{p:independent}, but follows
from the local observability principle.

\begin{table}
\tbl{Dimensionality identities implied by Postulates.}
{\begin{tabular}{|l|c|l|}
\toprule
$\Longrightarrow$ &  &\\ 
\hline\hline
Postulate 2 & $\dim(\Cntset_\Reals)=\adm(\Stset)+1$ & ($D_2$)\\
\hline
Postulate 3 & $\adm(\Stset_{12})=\adm(\Stset_1)\adm(\Stset_2)+\adm(\Stset_1)+\adm(\Stset_2)$ & ($D_3$)\\ 
\hline
Postulate 4 & $\adm(\Stset)=\idim{\Stset^{\times 2}}-1$  & ($D_4$)\\ 
\hline
($D_3$)+($D_4$) & $\adm(\Stset^{\times 2})=\idim{\Stset^{\times 2}}^2-1$  & ($D_{34}$)\\ 
\hline
($D_{34}$) & $\adm(\Stset)=\idim{\Stset}^2-1$  & ($D_{34}'$)$^{\text a}$\\ 
\hline
($D_4$+$D_{34}'$) & $\idim{\Stset^{\times 2}}=\idim{\Stset}^2$ & ($\otimes$) \\ 
\hline
Postulate 5 & $\adm(\Trnset)=\adm(\Stset^{\times 2})+1$  & ($\Trnset$)\\ 
\hline
($D_2$)+($D_{34}'$) & $\adm(\Cntset_\Reals)=\idim{\Stset}^2$  & ($\Cntset$)\\ 
\botrule
\end{tabular}}
\begin{tabnote}
$^{\text a}$ Generalizing from convex sets of states of bipartite systems to any convex set of states.\\
\end{tabnote}
\label{t:dimids}
\end{table}

Postulate \ref{p:Bell} now gives identity ($D_4$) in Table \ref{t:dimids}. By comparing this with the
affine dimension of the bipartite system, we get identity ($D_{34}$), and generalizing to any convex set
we get identity ($D_{34}'$) corresponding to the dimension of the quantum convex sets $\Stset$ due to the
underlying Hilbert space.  Moreover, upon substituting identity ($D_4$) one obtains identity
($\otimes$) which is the quantum product rule for informational dimensionalities corresponding to
the quantum {\em tensor product}. To summarize, it is worth noticing that the {\em quantum 
  dimensionality rules} ($D_3$) and ($\otimes$) follow from Postulates \ref{p:locobs} and \ref{p:Bell}.
Postulate \ref{p:faith}, on the other hand, implies identity ($\Trnset$).

\section{The complex Hilbert space structure for finite dimensions}
The faithful state $\Phi$ provides a symmetric bilinear form $\Phi(\cA,\cB)$ over $\Cntset_\Reals$,
from which one can extract a positive scalar product over $\Cntset_\Reals$ as $|\Phi|(\cA,\cB)$,
where $|\Phi|:=\Phi_+-\Phi_-$ is the absolute value of $\Phi$ (the absolute value can be defined
thanks to the fact that $\Phi$ is real symmetric, whence it can be diagonalized over
$\Cntset_Reals$). Upon denoting by $\map{P}_\pm$ the orthogonal projectors over the linear space
corresponding to positive and negative eigenvalues, respectively, one has
$|\Phi|(\cA,\cB)=\Phi(\cA,\varsigma(\cB))$, where $\varsigma(\cA):=(\map{P}_+-\map{P}_-)(\cA)$.  The
map $\varsigma$ is an involution, namely $\varsigma^2=\map{I}$. The fact that the state is also
preparationally faithful implies that the scalar product is strictly positive, namely
$|\Phi|(\cC,\cC)=0$ implies that $\cC=0$ (see Ref.~\refcite{darianoVax2006}). Now, being
$|\Phi|(\cA,\cB)$ a strictly positive real symmetric scalar product, the linear space
$\Cntset_\Reals$ of generalized effects becomes a \underline{real} pre-Hilbert space, which can be
completed to a Hilbert space in the norm topology. For finite dimensional convex set $\Stset$ one
has Eq. ($D_2$) in Table \ref{t:dimids}, which follows from the fact that since $\Cntset_\Reals$ is
just the space of the linear functionals over $\Stset$, it has an additional dimension corresponding
to normalization. But from Eq. ($D_2$) and ($D_{34}'$) one has identity ($\Cntset$), which implies
that $\Cntset_\Reals$ as a real Hilbert space is isomorphic to the real Hilbert space of Hermitian
complex matrices representing selfadjoint operators over a complex Hilbert space $\sH$ of dimensions
$\dim(\sH)=\idim{\Stset}$. This last assertion is indeed the Hilbert space formulation of Quantum
Mechanics, from which one can recover the full mathematical structure. In fact, once the generalized
effects are represented by Hermitian matrices, the physical effects will be represented as elements
of the truncated convex cone of positive matrices, the physical transformations will be represented
as CP identity-decreasing maps over effects, and finally, states will be represented as density
matrices via the Bush version \cite{busch} of the Gleason theorem, or via our state-effect
correspondence coming from the preparationally faithfulness of $\Phi$.

\section{Infinite dimension: the C${}^*$-algebra of transformations}
For infinite dimensions we cannot rely on the dimensionality identities in Table \ref{t:dimids}, and
we need an alternative way to derive Quantum Mechanics, such as the construction of a C$^*$-algebra
representation of generalized transformations. In order to do that we need to extend the real Banach
algebra $\Trnset_\Reals$ to a complex algebra, and for this we need to derive the {\em adjoint} of a
transformation from the five postulates (we will see that indeed only four of the five postulates
are needed). The adjoint is given as the composition of {\em transposition} and {\em
  complex-conjugation} of physical transformations, both maps being introduced operationally on the
basis of the existence of a symmetric dynamically faithful state due to Postulate \ref{p:faith}.
The {\em complex conjugate} map will be an extension to $\Trnset_\Reals$ of the involution
$\varsigma$ of Section \ref{s:finiteHilbert}. With such an adjoint one then derives a GNS
representation \cite{GelfandNeumark} for transformations, leading to a C${}^*$-algebra.
\paragraph{The transposed transformation.}
\par For a symmetric bipartite state that is faithful both dynamically and preparationally, for every
transformation on system 1 there always exists a (generalized) transformation on system 2 giving the
same operation on that state. This allows us to introduce operationally the notion of {\em
  transposed transformation} as follows.  For a {\em faithful} bipartite state $\Phi$, the {\em
  transposed transformation} $\tA'$ of the transformation $\tA$ is the generalized transformation
which when applied to the second component system gives the same conditioned state and with the same
probability as the transformation $\tA$ operating on the first system, namely $(\tA,\tI)\Phi=(\tI,\tA')\Phi$
\bigskip
\begin{figure}[h]
\begin{center}
\parbox{2.1in}{
\setlength{\unitlength}{800sp}
    \begin{picture}(8745,3219)(931,-3565)
      {\thicklines \put(5401,-1261){\oval(1756,1756)}}
      {\put(1801,-1261){\line(1, 0){2700}}}
      {\put(6301,-1261){\vector(1, 0){3300}}}
      {\put(1801,-3361){\vector(1, 0){7800}}}
      \put(2026,-2611){\makebox(0,0)[b]{$\Phi$}}
      \put(5401,-1486){\makebox(0,0)[b]{$\tA$}}
      \put(9676,-2811){\makebox(0,0)[b]{$(\tA,\tI)\Phi$}}
    \end{picture}
}
  \hspace*{0pt}
  \parbox{2.1in}{
    \setlength{\unitlength}{800sp}
    \begin{picture}(8745,3219)(931,-3565)
      {\thicklines \put(5401,-3461){\oval(1756,1756)}}
      {\put(1801,-3361){\line(1, 0){2700}}}
      {\put(6301,-3361){\vector(1, 0){3300}}}
      {\put(1801,-1261){\vector(1, 0){7800}}}
      \put(2026,-2611){\makebox(0,0)[b]{$\Phi$}}
      \put(5401,-3686){\makebox(0,0)[b]{$\tA'$}}
      \put(11676,-2811){\makebox(0,0)[b]{$(\tI,\tA')\Phi\equiv(\tA,\tI)\Phi$}}
    \end{picture}
}
\bigskip
\caption{Illustration of the operational concept of {\em transposed transformation}.}
\end{center}
\end{figure}
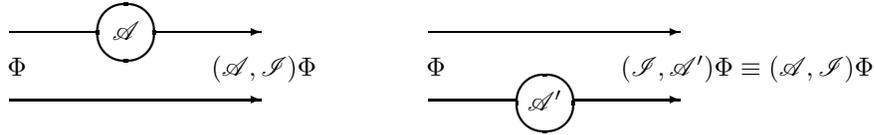
or, equivalently $\Phi(\cB\circ\tA,\cC)=\Phi(\cB,\cC\circ\tA')$ $\forall \cB,\cC\in\Cntset$.

It is easy to check the axioms of transposition ($(\tA+\tB)'= \tA'+\tB'$, $(\tA')'=\tA$, $(\tA\circ\tB)'=
\tB'\circ\tA'$) and that $\tI'=\tI.$ Unicity is implied by faithfulness.
\paragraph{The complex conjugated transformation.}
Due to the presence of the involution $\varsigma$, the transposition $\tA\to\tA'$ does not work as
an adjoint for the scalar product $|\Phi|(\cA,\cB)$ (it works as an adjoint for the symmetric
bilinear form $\Phi$, which is not positive).  In order to introduce an adjoint for generalized
transformations (with respect to the scalar product between effects) one needs to extend the
involution $\varsigma$ to generalized transformations. With a procedure analogous to that used for
effects we introduce the absolute value $|\Phi|$ of the symmetric bilinear form $\Phi$ over
$\Trnset_\Reals$, whence extend the scalar product to $\Trnset_\Reals$.  Clearly, since the bilinear
form $\Phi(\tA,\tB)$ will anyway depend only on the informational equivalence classes $\cA$ and
$\cB$ of the two transformations, we have many extensions of $\varsigma$ which work equally well.
Upon defining $\tA^\varsigma:=\varsigma(\tA)$, one has $\tA^\varsigma\in\varsigma(\cA)$, and clearly
one has $\varsigma^2(\tA)=\varsigma(\tA^\varsigma)\in\cA$, but generally $\varsigma^2(\tA)\neq\tA$.
However, one can always consistently choose the extension such that $\varsigma^2(\tA)=\tA$.  The
idea is now that such an involution plays the role of the {\em complex conjugation}, such that the
composition of $\varsigma$ with the transposition provides the adjoint.
\smallskip
\paragraph{The adjoint transformation.}
Due to the fact that transformations act on effects from the right---\ie
$\cB\circ\tA\in\underline{\tB\circ\tA}$---in order to keep the usual action on the left in the
representation of transformations over generalized effects it is convenient to redefine the scalar
product via the bilinear form $\Phi(\tA',\tB')$ over transposed transformations.  Therefore, we
define the scalar product between generalized effects as follows
\begin{equation}\label{scalproddef}
{}_\Phi\!\<\cB|\cA\>_\Phi:=\Phi(\cB',\varsigma(\cA')).
\end{equation}
Notice how in this way one recovers the customary operator-like action of transformations from the
left $|\underline{\tC\circ\tA}\>_\Phi= |\tC\circ\cA\>_\Phi$ which follows from
${}_\Phi\!\<\tC\circ\cA|\tB\>_\Phi=\Phi(\cA'\circ\tC',\varsigma(\cB'))$. In the following we will
equivalently write the entries of the scalar product as generalized transformations or as
generalized effects, with ${}_\Phi\!\<\tA|\tB\>_\Phi:={}_\Phi\!\<\cA|\cB\>_\Phi$, the generalized
effects being the actual vectors of the linear factor space of generalized transformations modulo
informational equivalence.

For {\em composition-preserving} involution (\ie
$\varsigma(\tB\circ\tA)=\tB^\varsigma\circ\tA^\varsigma$) one can easily verify\cite{darianoVax2006}
that $\tA^\dag:=\varsigma(\tA')$ works as an adjoint for the scalar product, namely
\begin{equation}
{}_\Phi\!\<\tC^\dag\circ\cA|\cB\>_\Phi={}_\Phi\!\<\cA|\tC\circ\cB\>_\Phi.
\end{equation}
In terms of the adjoint the scalar product can also be written as
${}_\Phi\!\<\tB|\tA\>_\Phi=\Phi|_2(\tA^\dag\circ\tB)$.  The involution $\varsigma$ is
composition-preserving if $\varsigma(\Trnset)=\Trnset$ namely if the involution preserves physical
transformations.  Indeed, for such an involution one can consider its action on transformations
induced by the involutive isomorphism $\omega\to\omega^\varsigma$ of the convex set of states
$\Stset$ defined as $\omega^\varsigma(\tA):=\omega(\varsigma(\tA))$,
$\forall\omega\in\Stset,\;\forall\tA\in\Trnset$. Consistency with state-reduction
$\omega_\tA^\varsigma(\tB)\equiv\omega_{\tA^\varsigma}(\tB^\varsigma)$
$\forall\omega\in\Stset,\;\forall\tA,\tB\in\Trnset$ is then equivalent to
$\omega(\varsigma(\tB\circ\tA))=\omega(\tB^\varsigma\circ\tA^\varsigma)$
$\forall\omega\in\Stset,\;\forall\tA,\tB\in\Trnset$. The involution $\varsigma$ of $\Stset$ is just
the inversion of the principal axes corresponding to negative eigenvalues of the symmetric bilinear
form $\Phi$ of the faithful state.

\paragraph{The GNS construction and the C${}^*$-algebra.} By taking complex linear combinations of
generalized transformations and defining $\varsigma(c\tA)=c^*\varsigma(\tA)$ for $c\in\Cmplx$, we
can now extend the adjoint to complex linear combinations of generalized transformations, whose
linear space will be denote by $\Trnset_\Cmplx$. On the other hand, we can trivially extend the real
pre-Hilbert space of generalized effects $\Cntset_\Reals$ to a complex pre-Hilbert space
$\Cntset_\Cmplx$ by just considering complex linear combinations of generalized effects. The complex
algebra $\Trnset_\Cmplx$ (that we will also denote by $\aA$) is now a complex Banach algebra 
of transformations on the Banach space $\Cntset_\Cmplx$. We have now a scalar product
${}_\Phi\!\<\tA|\tB\>_\Phi$ between transformations, and an adjoint of transformations with respect
to such scalar product. Symmetry and positivity imply the bounding\cite{darianoVax2006}
${}_\Phi\!\<\tA|\tB\>_\Phi\leq\n{\tA}_\Phi\n{\tB}_\Phi$,
where we introduced the norm induced by the scalar product
$\n{\tA}_\Phi^2\doteq{}_\Phi\!\<\tA|\tA\>_\Phi$.
From the bounding for the scalar product it follows that the set $\aI\subseteq\aA$ of zero norm
elements $\tX\in\aA$ is a left ideal, \ie it is a linear subspace of $\aA$ which is stable under
multiplication by any element of $\aA$ on the left (\ie $\tX\in\aI$, $\tA\in\aA$ implies
$\tA\circ\tX\in\aI$).  The set of equivalence classes $\aA/\aI$ thus becomes a complex pre-Hilbert
space equipped with a symmetric scalar product. On the other hand, since the scalar product is
strictly positive over generalized effects, the elements of $\aA/\aI$ are indeed the generalized
effects, \ie $\aA/\aI\simeq\Cntset_\Cmplx$ as linear spaces. Therefore, informationally equivalent
transformations $\tA$ and $\tB$ correspond to the same vector, and there exists a generalized
transformation $\tX$ with $\n{\tX}_\Phi=0$ such that $\tA=\tB+\tX$, and $\n{\cdot}_\Phi$, which is a
norm on $\Cntset_\Cmplx$, will be just a semi-norm on $\aA$.  We can re-define anyway the norm on
transformations as $\n{\tA}_\Phi:=\sup_{\cB\in\Cntset_\Cmplx,\n{\cB}_\Phi\leq
  1}\n{\tA\circ\cB}_\Phi$. Completion of $\aA/\aI\simeq\Cntset_\Cmplx$ in the norm topology will
give a Hilbert space that we will denote by $\sH_\Phi$.  Such completion also implies that
$\Trnset_\Cmplx\simeq\aA$ is a complex C$^*$-algebra (\ie satisfying the identity
$\n{\tA^\dag\circ\tA}=\n{\tA}^2$), as it can be easily proved by standard
techniques\cite{darianoVax2006}. The fact that $\aA$ is a C${}^*$-algebra---whence a Banach
algebra---also implies that the domain of definition of $\pi_\Phi(\tA)$ can be easily extended to
the whole $\sH_\Phi$ by continuity.

The product in $\aA$ defines the action of $\aA$ on the vectors in $\aA/\aI$, by associating to each
element $\tA\in\aA$ the linear operator $\pi_\Phi(\tA)$ defined on the dense domain
$\aA/\aI\subseteq\sH_\Phi$ as follows
\begin{equation}
\pi_\Phi(\tA)|\cB\>_\Phi\doteq|\underline{\tA\circ\tB}\>_\Phi.
\end{equation}
\paragraph{Born rule.} 
From the definition (\ref{scalproddef}) of the scalar product the Born rule rewrites in terms of the pairing
$\omega(\cA)={}_\Phi\<\pi_\Phi(\cA)|\pi_\Phi(\omega)\>_\Phi$, 
with representations of states
$\pi_\Phi(\omega)=\widetilde\cT_\omega:=\cT_\omega'/\Phi(\tI,\cT_\omega)$, and of effects 
$\pi_\Phi(\cA)=\cA'$ (see Ref.~\refcite{darianoVax2006}). 
Then, the representation of transformations is 
$\omega(\cB\circ\tA)={}_\Phi\<\cB'|\pi_\Phi(\tA^\varsigma)|\pi_\Phi(\omega)\>_\Phi$.

\end{document}